\documentclass[12pt]{article}
\usepackage{a4wide}
\usepackage{amssymb}
\usepackage{mathtools}
\begin{document}
{\renewcommand{\thefootnote}{\fnsymbol{footnote}}
\begin{center}
{\LARGE  Spherically symmetric canonical quantum gravity}\\
\vspace{1.5em}
Suddhasattwa Brahma\footnote{e-mail address: {\tt sxb1012@psu.edu}}
\\
\vspace{0.5em}
Institute for Gravitation and the Cosmos,\\
The Pennsylvania State
University,\\
104 Davey Lab, University Park, PA 16802, USA\\
\vspace{1.5em}
\end{center}
}

\setcounter{footnote}{0}

\newcommand{\nn}{\nonumber}
\newcommand{\be}{\begin{equation}}
\newcommand{\ee}{\end{equation}}
\newcommand{\bea}{\begin{eqnarray}}
\newcommand{\eea}{\end{eqnarray}}

\begin{abstract}
Canonical quantization of spherically symmetric space-times is carried out, using real-valued densitized triads and extrinsic curvature components, with specific factor ordering choices ensuring in an anomaly free quantum constraint algebra. Comparison with previous work \cite{Thiemann:1992jj} reveals that the resulting physical Hilbert space has the same form, although the basic canonical variables are different in the two approaches. As an extension, holonomy modifications from Loop Quantum Gravity are shown to deform the Dirac space-time algebra, while going beyond `effective' calculations.
\end{abstract}



\section{Introduction}
It has been shown that spherically symmetric gravity can be canonically quantized in a nonperturbative manner  \cite{Thiemann:1992jj, Kuchar:1994zk}.  In the original work by Kastrup and Thiemann, the theory was reformulated in terms of self-dual Ashtekar variables, i.e. self-dual connection coefficients and real densitized triads were used as the canonical variables to quantize the system. However, it has since been customary to introduce a real valued Immirzi parameter to define an Ashtekar-Barbero connection for several technical reasons \cite{9780511755682, 9780511755804}.\footnote{This helps in setting up a well-defined Hilbert space that does not have non-compact $sl(2,\mathbb{C})$ holonomies for complex connections. Additionally we do not have to worry about solving complicated reality conditions.} In the first part of this article, we intend to carry out a Wheeler-Dewitt quantization of spherically symmetric field configurations where the basic variables are all real. The densitized triads are now conjugate to extrinsic curvature components, which are themselves real variables. We also eliminate one canonical pair of variables as compared to \cite{Thiemann:1992jj} by classically solving for the Gauss constraint and the  corresponding gauge flows it generates. Although the resulting \textit{classical} systems we end up with in the two cases are canonical transformations of each other, the quantum theories are based on different basic canonical variables, and consequently different equal time commutation relations. This ensures that the form of the gravitational constraints are also vastly different. Indeed the number of constraints per spatial point in \cite{Thiemann:1992jj} was three compared to the two we are left with in this case. Thus the system after quantization can end up being inequivalent in the two cases. Remarkably, as we shall show in this paper, the physical Hilbert space for both these systems end up having the same form in the triad representation. This result goes to demonstrate the robustness of the canonical quantization procedure for such a system, irrespective of several details in the two methods.

In order to solve for this midi-superspace model, we consider the classical phase space as given in \cite{9780511921759, JR, Bojowald:2004af, Bojowald:2005cb}. We show that there exists particular factor ordering choices, in the formal sense, for the Hamiltonian and diffeomorphism constraints such that the quantum version of the Dirac space-time algebra is well defined. This is a non trivial task since we do not have a Lie algebra in this case and thus must be careful to ensure that the structure functions appearing on the right hand side of the algebra appears to the left of the constraints, to ensure the Dirac consistency condition. This has been elaborated on later in Sec(\ref{Algebra}). Using one such factor-ordering for the constraints, we employ Dirac quantization for extracting the physical Hilbert space and finally construct a suitable inner product on it. The basis for $\mathcal{H}_{\text{physical}}$, in the triad representation, turns out to be exactly the same for us as in \cite{Thiemann:1992jj}, although the technical details of the two approaches are quite different. However, then we do choose our physical states in a manner analogous to that in \cite{Thiemann:1992jj} in order to induce an inner product on this physical Hilbert space.

In the last section, we show the effects of including holonomy corrections from Loop Quantum Gravity on spherically symmetric space-times. In LQG, there is no well-defined operator corresponding to the connection components \cite{9780511755682, 9780511755804, Ashtekar:2004eh}. Instead the constraints are modified such that they are represented by holonomy variables. In this framework, quantum correction functions are introduced to capture the effects of replacing connections by their corresponding holonomies. Once again, we give a consistent operator ordering choice for the constraint operators that keeps the quantum Dirac space-time algebra anomaly-free. Although the algebra remains first-class even in the presence of such correction functions, the structure functions are shown to get deformed by additional phase-space functions in their presence. Since these structure functions also encode the background structure of space-time itself, this indicates that classical background structures may not be valid anymore in the presence of such deformations. Our results seem to be consistent with expectations from `effective' theories  \cite{Bojowald:2011aa, Bojowald:2012qya, Bojowald:2014jwa}, although they go beyond them by further restricting the exact form of gravitational constraints. The consequences of having such deformations of the algebra is also briefly discussed.

\section{Classical phase space}
In canonical general relativity (GR), the hamiltonian is a linear sum of constraints and thus trivial on the constraint surface. Since the theory is diffeomorphism invariant, space-time diffeomorphisms can be realized as gauge transformations on phase space functions generated by first-class constraints. As a consequence, time evolution is also a pure gauge transformation. The canonical variables of the theory, in this first-class formalism, are chosen to be the $su(2)$-valued Ashtekar-Barbero connection $A_a^i$ and the densitized triad vector fields $E^a_i$, both of which are functions on the three dimensional manifold $\Sigma$. The spatial metric on the manifold $\Sigma$ can be written in terms of the triad, $q q^{ab}=E^a_iE^b_j\delta^{ij}$ with $q:=\text{det}(q_{ab})$, whereas the Ashtekar-Barbero connection is related to the extrinsic curvature of $\Sigma$ and the triad-compatible spin connection $\Gamma_a^i$ according to the equation $A_a^i=\Gamma_a^i+\gamma K_a^i$, where $\gamma$ is a real constant called the Immirzi parameter\footnote{This parameter plays no role in the classical theory as it can be changed by canonical transformations.} and $K_a^i:=(\text{det}E)^{-1/2}K_{ab} E^{bi}$. The action for GR, in terms of these variables, is given by

\bea\label{FullAction}
S_{\text{4D}}=\int\mathrm{d}t\left(\frac{1}{8\pi G\gamma}\int_{\Sigma}\mathrm{d}^3x E^a_i\dot{A}_a^i-\int_{\Sigma}\mathrm{d}^3x\left[\lambda^i\mathcal{G}_i +N^a\mathcal{D}_a+ N\mathcal{H}\right] \right),
\eea
where $G$ is Newton's constant in four dimensions, and $\lambda^i, N^a$ and $N$ are Lagrange multipliers for the various first-class constraints. The Gauss constraint, $G[\lambda^i]=\frac{1}{8\pi G\gamma}\int \mathrm{d}^3x \lambda^i\mathcal{G}_i$, generates $SU(2)$ transformations, the diffeomorphism constraint, $D[N^a]=\frac{1}{8\pi G\gamma}\int \mathrm{d}^3x N^a\mathcal{D}_a$, generates spatial diffeomorphisms and the Hamiltonian constraint, $H[N]=\frac{1}{8\pi G\gamma}\int \mathrm{d}^3x N\mathcal{H}$, generates time evolution. It is now easy to identify $N^a$ and $N$ as the familiar shift vector and lapse function respectively, as in the ADM formulation of GR \cite{Arnowitt:1959ah}. These constraints can all be expressed explicitly in terms of the Ashtekar variables.

Details of the symmetry reduction for spherical symmetry with Ashtekar variables had been carried out in several previous work \cite{9780511921759, JR, Bojowald:2004af, Bojowald:2005cb}. In this article, we shall closely follow the conventions and notations as in \cite{JR}. After spherical symmetry reduction, we have three independent canonical pairs given by
\bea\label{CanPairs}
\left\{A_x(x),\frac{1}{2\gamma}{{E^x}}(y)\right\}&=&G\delta(x-y),\\
\left\{{K_\phi}(x),{E^{\phi}}(y)\right\}&=&G\delta(x-y),\\
\left\{\eta(x),\frac{1}{2\gamma}P^{\eta}(y)\right\}&=&G\delta(x-y).
\eea
Here $x$ and $y$ parameterizes the radial coordinate and we have a one dimensional Dirac delta function on the RHS of the above equations. In terms of these variables, the spatial metric on the three dimensional manifold can be expressed as
\bea\label{spatmet}
\mathrm{d}q^2=\frac{({E^{\phi}})^2}{|{{E^x}}|}\mathrm{d}x^2+|{{E^x}}|
\mathrm{d}\Omega^2,
\eea
where the usual angular part is given by $\mathrm{d}\Omega^2=\mathrm{d}\theta^2+\sin^2\theta\mathrm{d}\phi^2$.
The symmetry reduced action looks like
\bea\label{Action}
S&=&\int\mathrm{d}t \Bigg[\frac{1}{2G\gamma}\int \mathrm{d}x\left({{E^x}}\dot{A}_x + 2\gamma {E^{\phi}}\dot{K}_{\phi}+ P^{\eta}\dot{\eta}\right)\nonumber\\
&&\,\,\,\,\,\,\,\,\,\,\,\,\,\,\,\,\,\,\,\,\,\,\,-\int \mathrm{d}x \left(\lambda\mathcal{G}+N^x\mathcal{D}+N\mathcal{H}\right) \Bigg],
\eea
where the various constraints are as follows. The only non-trivial component of the Gauss constraint generating $U(1)$-gauge transformations is
\bea\label{Gauss}
G[\lambda]=\frac{1}{2G\gamma}\int\mathrm{d}x\,\,\lambda \left({{E^{x}}}'+ P^{\eta}\right).
\eea
The prime denotes derivative with respect to the radial coordinate $x$.
Similarly, the diffeomorphism constraint also has a single non-trivial component that generates spatial diffeomorphisms in the radial direction
\bea\label{diffeo1}
D[N^x]=\frac{1}{2G}\int\mathrm{d}x \,\,N^x\left(2{E^{\phi}}{K_\phi}'-\frac{1}{\gamma}A_x{{E^{x}}}' +\frac{1}{\gamma}\eta'P^{\eta}\right).
\eea
Finally time evolution is generated by the Hamiltonian constraint
\bea\label{Ham}
H[N]=\frac{-1}{2G}\int\mathrm{d}x\,\, N |{{E^x}}|^{-1/2}\left({K_\phi}^2{E^{\phi}}+2{K_\phi}K_x{{E^x}}+
[1-\Gamma_{\phi}^2]{E^{\phi}}+2\Gamma_{\phi}'{{E^x}}\right),
\eea
where $\Gamma_{\phi}=-{{E^x}}'/(2{E^{\phi}})$ and $K_x=\frac{1}{\gamma}\left(A_x+\eta'\right)$.

In what follows, we are going to assume that we have solved the Gauss constraint (\ref{Gauss}), and thus eliminated the canonical pair $(\eta,P^{\eta})$. We are left with two constraints per space-time point, and the two canonical pairs $({{E^x}}, K_x), ({E^{\phi}}, {K_\phi})$. Thus one of the terms drops out to simplify the diffeomorphism constraint as
\bea\label{diffeo}
D[N^x]=\frac{1}{2G}\int\mathrm{d}x \,\,N^x\left(2{E^{\phi}}{K_\phi}'-K_x{E^{x}}'\right).
\eea
The modified symplectic structure is
\bea\label{PB}
\left\{K_x(x),{{E^x}}(y)\right\}&=& 2 G\delta(x-y),\\
\left\{{K_\phi}(x),{E^{\phi}}(y)\right\}&=& G\delta(x-y),
\eea
and all other Poisson brackets are equal to zero. Therefore, we are left with two gravitational constraints per spatial point and two canonical pairs, indicating that our model has no local physical degrees of freedom, as expected in spherically symmetric gravity. All the basic variables in our model, are thus, real-valued.
At this point, comparison with the canonical variables and the three gravitational constraints in \cite{Thiemann:1992jj} is helpful to realize how vastly distinct they are in form to that of ours.

\section{Quantization}

\subsection{Algebra $\mathcal{A}$}
We define a $*-$algebra $\mathcal{A}$ by converting the Poisson brackets into equal time canonical commutation relations. The hats are to remind us that these are quantum operators corresponding to the classical variables. Only the non-trivial commutators are shown below.
\bea\label{alg}
\left[\hat{K}_x(x),\hat{E}^x(y)\right]&=& 2i\hbar G\delta(x-y),\\
\left[\hat{K}_{\phi}(x),\hat{E}^{\phi}(y)\right]&=& i\hbar G\delta(x-y).
\eea
(We note here that the above definitions for the algebra are given in a formal sense. To be more rigorous, we should smear the canonical variables with smooth functions to get rid of the delta functions on the RHS. However, from our viewpoint, we do not fully regularize the constraints but rather pay attention to factor ordering details for getting a consistent  Dirac space-time algebra. This point shall be elaborated later on.)

These real Ashtekar variables must also satisfy the $*-$relations
\bea
(\hat{E}^x(x))^*=\hat{E}^x(x),\,\,\,\,\, (\hat{E}^{\phi}(x))^*=\hat{E}^{\phi}(x)\label{star1},\\
(\hat{K}_{x}(x))^*=\hat{K}_{x}(x),\,\,\,\,\, (\hat{K}_{\phi}(x))^*=\hat{K}_{\phi}(x)\label{star2}.
\eea
We shall use these relations (\ref{star1}, \ref{star2}) to determine the physical scalar product as in, say, \cite{Ashtekar:1994kv, Tate:1992hc, Rendall:1993jh, Rendall:1994yq}. These $*-$relations on the basic variables are to be imposed as adjointness conditions with respect to the inner product, as we shall explore in detail later on.

\subsection{Triad Representation}
We next need to construct a representation of the above $*-$algebra $\mathcal{A}$ on some linear space. In this article, we choose the `triad representation', whereby  the basic variables act via linear operators as
\bea\label{reprtn}
\hat{E}^x(x)\Psi= {{E^x}}(x)\Psi, \,\,\,\,\, \hat{E}^\phi(x)\Psi= {{E^\phi}}(x)\Psi,\\
\hat{K}_x(x)\Psi=2 \left(Gi\hbar\right)\frac{\delta}{\delta {{E^x}}(x)}\Psi, \,\,\,\,\, \hat{K}_\phi(x)\Psi= \left(Gi\hbar\right)\frac{\delta}{\delta {E^\phi}(x)}\Psi,
\eea
where $\Psi=\Psi[{{E^x}}, {E^{\phi}}]$ is a smooth (differentiable) functional of the triad variables, which remains to be determined.

\section{Analysis of the constraint algebra}\label{Algebra}
The diffeomorphism and Hamiltonian constraints for our model are given in (\ref{diffeo}, \ref{Ham}). Classically, these constraints satisfy Dirac's space-time algebra \cite{Dirac:1958sc}
\bea
 \left\{D[N^x], D[M^x]\right\} &=& D[\mathcal{L}_{N^x}M^x],\\
 \left\{H[N], D[N^x]\right\} &=& -H[\mathcal{L}_{N^x}N],\\
 \left\{H[N], H[M]\right\} &=& D[(NM'-MN')|{{E^x}}|({E^{\phi}})^{-2}].\label{HypersurfaceDefnAlg}
\eea
Our goal is to make sure the quantum constraint operators corresponding to the classical constraints obey the so-called `Dirac consistency' condition. What this means is as follows. Schematically both the gravitational constraints can be written as $C_I$ (with $I=1,2$ for the diffeomorphism and Hamiltonian constraint respectively). Classically the constraints must be satisfied ($C_I=0$) to ensure that the resulting system is space-time covariant, even though we started with slicing up space and time in the canonical formulation. In the quantum theory, the presence of these constraint operators imply that not all wave-functions of the form $\Psi[{{E^x}}, {E^{\phi}}]$ correspond to valid physical states. Dirac quantization implies that $\Psi_{\text{phys}}$ is a physical state only if it is annihilated by both the gravitational constraints \cite{dirac2001lectures},
\bea\label{DiracConstEqn}
\hat{C}_I \Psi_{\text{phys}}=0.
\eea
Obviously this implies that both the constraints acting consecutively on $\Psi_{\text{phys}}$ gives zero and thus their commutator must also annihilate the physical state.
\bea\label{CommAnnhi}
\left[\hat{C}_I, \hat{C}_J\right] \Psi_{\text{phys}}=0.
\eea
Since the above relation must hold for any arbitrary physical state, we must have that the commutator of two constraints must itself be a linear combination of constraints. This is the requirement for the constraints to be `first-class' in the language of Dirac. For the quantum constraint operators, this means
\bea\label{CommConst}
\left[\hat{C}_I, \hat{C}_J\right]=\hat{f}_{IJK}\hat{C}_K.
\eea
The coefficients on the RHS can be operators themselves depending on phase space variables. In the case of our model, the only non-trivial structure function showing up in the $[H,H]$ bracket is $|\hat{E}^x|(\hat{E}^{\phi})^{-2}$, as is evident from (\ref{HypersurfaceDefnAlg}). However, the important thing is that this structure function appears to the left of the constraint operator appearing on the RHS. This must be so in order to have the commutator annihilate a physical state. This is the essence of the `Dirac consistency' condition. Failing to satisfy this condition implies that the only physical state satisfying equations (\ref{DiracConstEqn}) is the trivial one. Unless we can find a factor ordering choice for the quantum constraint operators such that `Dirac consistency' condition is satisfied, there shall be gauge anomalies arising due to the quantization procedure.

As we shall demonstrate next, there exist consistent factor ordering choices, for which the quantum algebra remains first-class. Here we shall refer to ordering of the constraint operators in a formal sense. For a full quantum analysis, we should regularize the quantum constraint operators first and then look for a consistent factor-ordering choice\footnote{We wish to thank Casey Tomlin for pointing out this article to us.} \cite{Tsamis:1987wf}. Although we never need to utilize distributional relations of the form $f(y)g(x)\delta'(x-y)=f(x)g(x)\delta'(x-y) + f'(x)g(x)\delta(x-y)$, which was shown to be cause of several ambiguities in \cite{Tsamis:1987wf}, we need to make sense of derivatives of delta-functions that appear in the calculations. (This, of course, stems from the fact that we have several operators defined at the same spatial point in these constraints.)  We can take the point of view that the constraint operators have been regularized, say, by point-splitting. In this case, instead of having two operators with the same argument $x$, we replace one of them by a dummy argument $y$ and then multiply the operator-product by some smearing function very sharply peaked around $x=y$, while  integrating over the dummy variable. In the limit of the smearing function approaching the delta function $\delta(x-y)$, we encounter divergences which are then removed by a rigorous subtraction scheme. Instead of doing these explicit calculations, we carry out the formal manipulations to show that the particular operator ordering of the constraints leads to a first-class algebra. Then we shall also have to show that the physical wave-functionals are annihilated by this particular choice of the factor-ordered constraint operators to get the physical Hilbert space.

Alternatively, we can take the point of view that this is a gauge theory defined on some finite lattice, whereby the constraint operators are already regularized by some lattice parameter. We can then carry out the formal manipulations before removing the lattice regulator at the end, at which point we shall, once again, require a precise subtraction to define how regularization works in this scheme. But our main interest lies in going beyond `effective' models to show that there are further restrictions on how the quantum constraint operators must be ordered so as to have a well-defined constraint algebra. Thus we can ignore the details of the regularization scheme for our purposes.

\subsection{Factor ordering choices}
In order to demonstrate that a consistent factor ordering choice exists, we shall adopt the plan of starting with an ansatz and then showing that such an ordering works. We shall show that there are (at least) two different choices for the gravitational constraints that seem to work. As far as the Hamiltonian constraint goes, only the first two terms have both triads and connection components in them, thereby creating an ambiguity in their ordering. The rest of the terms only consists of triads and thus they seem to be free from factor ordering choices. For the diffeomorphism constraint, we need to pick an operator ordering for both the terms.

Let us start by factor ordering the Hamiltonian constraint. We shall drop all the hats on the quantum operators from now on, which were introduced to differentiate them from their classical counterparts (but remain careful with orderings).
\bea\label{HamQ}
H[N]&=&\frac{-1}{2G}\int \mathrm{d}x \,\,N\Bigg[({{E^x}})^{-1/2}{E^{\phi}}{K_\phi}^2 + 2({{E^x}})^{1/2}K_{x}{K_\phi} +({{E^x}})^{-1/2}{E^{\phi}}  \\
& & - \frac{1}{4}({{E^x}})^{-1/2}(E^{x\prime})^{2}({E^{\phi}})^{-1} - ({{E^x}})^{1/2}E^{x\prime\prime}({E^{\phi}})^{-1} +({{E^x}})^{1/2}E^{x\prime}({E^{\phi}})^{-2}E^{\phi\prime}\Bigg].\nonumber
\eea
We shall name this ordering choice for the Hamiltonian constraint the `normal ordering' choice since in this case the triads are pushed to the left and the conjugate momenta to the right. If we start with this ordering for the Hamiltonian constraint and calculate the $[H,H]$ bracket, then the factor ordering for the diffeomorphism constraint comes out to be
\bea\label{diffeoQ}
D[N^x]=\frac{1}{2G}\int\mathrm{d}x \,\,N^x\left[2{E^\phi} K'_\phi-K_xE^{x\prime}\right].
\eea
The details of the calculations which show that this is a consistent factor-ordering choice for both the gravitational constraints are shown in Appendix A. It is interesting to note that the diffeomorphism constraint is \textit{not} `normal-ordered' in the sense of the Hamiltonian constraint, a requirement from the closure of the algebra. This form of the diffeomorphism constraint (where the relative positions of $K_x$ and $E^x$ are different from that of $K_\phi$ and $E^\phi$) implies that when we go on to solve for the physical states we shall need to regularize the constraints by point-splitting, such that the operator generates infinitesimal diffeomorphisms. This is consistent with our expectation from before about regularising the constraint operators. This shall be implicitly assumed when we apply the Dirac formalism to obtain the physical Hilbert space in the next section.

It is obvious from this factor ordering choice that there exists another similar ordering choice where for the Hamiltonian constraint operator, all the triads are pushed to the right whereas the curvature components are moved to the left. This leads to a diffeomorphism constraint operator which is similar to the one above in (\ref{diffeoQ}), with the canonical variables commuting places with each other. However, we do not consider this to be a new ordering choice, since this is exactly the opposite of our `normal-ordering' choice.

There exists, at least, one other consistent factor-ordering choice as follows. In this case, the Hamiltonian constraint gets an ordering of the form
\bea\label{HamQ1}
H[N]&=&\frac{-1}{2G}\int \mathrm{d}x \,\,N\Bigg[({{E^x}})^{-1/2}{K_\phi}^2{E^{\phi}} + 2({{E^x}})^{1/2}K_{x}({E^{\phi}})^{-1}{K_\phi}{E^{\phi}} +({{E^x}})^{-1/2}{E^{\phi}}  \\
& & - \frac{1}{4}({{E^x}})^{-1/2}(E^{x\prime})^{2}({E^{\phi}})^{-1} - ({{E^x}})^{1/2}E^{x\prime\prime}({E^{\phi}})^{-1} +({{E^x}})^{1/2}E^{x\prime}({E^{\phi}})^{-2}E^{\phi\prime}\Bigg].\nonumber
\eea
In this case, the diffeomorphism constraint has to take the form
\bea\label{diffeoQ1}
D[N^x]=\frac{1}{2G}\int\mathrm{d}x \,\,N^x\left[2K'_\phi {E^\phi} - E^{x\prime}K_x\right].
\eea
Although we do not show the explicit calculations that this is indeed a consistent factor ordering choice for the gravitational constraints, they go similar to the calculations shown in Appendix A for the `normal ordering' case. At this point, we are unable to choose between either of these ordering choices, and one of them is as good as the other. However, in the next section, we shall demonstrate that only one of these ordering choices gives us a non-trivial physical state as the solution to the constraint operators. It is also important to mention here that there certainly are many different factor ordering choices that do \textit{not} obey the Dirac consistency relation, for instance a totally symmetric (or Weyl) ordering for each of the terms of the Hamiltonian constraint.

\section{The physical Hilbert space}
The Dirac quantization condition \cite{dirac2001lectures} tells us that the physical subspace of states must be annihilated by the gravitational constraints
\bea\label{Dirac1}
\mathcal{D}\Psi_{\text{phys}}=0,\\
\mathcal{H}\Psi_{\text{phys}}=0,
\eea
where $\mathcal{D}$ and $\mathcal{H}$ are the unsmeared version of the constraints. We can treat these equations (\ref{Dirac1}) as functional differential equations and look for solutions which would correspond to the physical wave-functions. A theorem, proved in \cite{Thiemann:1992jj}, states that such solutions are guaranteed to exist, at least locally, if the Dirac space-time algebra remains first-class, and if the number of constraints per spatial point coincides with the number of configuration space variables (two for our model). In order to solve these equations, we first need to pick one of the consistent ordering choices mentioned in the section above. We pick the `normal ordering' choice for the constraints to begin with.

The diffeomorphism constraint, as a functional differential equation, takes the form
\bea\label{DiffeoEqn}
{E^\phi}(x)\frac{d}{dx}\left[\frac{\delta\Psi}{\delta {E^\phi}(x)}\right] = \frac{\delta}{\delta {{E^x}}(x)}\left[{{E^x}}'(x)\Psi\right].
\eea
The form of (\ref{DiffeoEqn}) shows that the wave-functional cannot be a local function of the spatial coordinate. The LHS of the above equation has functional derivatives w.r.t. ${E^\phi}(x)$, followed by a spatial derivative operator. However the RHS has a functional derivative w.r.t. ${{E^x}}(x)$ acting on the product of the wave-functional with the spatial derivative of ${{E^x}}(x)$. Thus the physical states can only be a functional of integrated out triads. When trying different ansatze, it is important to remember that the density weight of ${E^\phi}(x)$ is one while that of ${{E^x}}(x)$ is zero. Thus the functionals must be constructed with the integrand having the proper density weight.
Thus the solutions of (\ref{DiffeoEqn}) are of the form
\bea\label{DiffeEqnSoln}
\Psi&=&\Psi\left[\int\mathrm{d}x\, {E^\phi}(x)\,\,f\left[{{E^x}}(x),  \left(\frac{{{E^x}}'(x)}{{E^\phi}(x)}\right)\right]\right]  \nonumber\\
&=&\Psi\left[\int\mathrm{d}x\, {{E^x}}'(x)\,\, g\left[{{E^x}}(x),  \left(\frac{{{E^x}}'(x)}{{E^\phi}(x)}\right)\right]\right],\,\, \ldots
\eea
where $f$ stands for a yet undetermined functional. There are obviously infinite such solutions for the diffeomorphism constraint, which are represented by the dots above. Next we should plug these solutions in the equation for the Hamiltonian constraint, which  written in terms of functional derivative operators, (for simplicity we have set $G=\hbar=1$ in the following)
\bea\label{HamEqn}
& &-\left[({E^\phi}(x))^2\frac{\delta^2}{\delta({E^\phi}(x))^2} -4{{E^x}}(x){E^\phi}(x)\frac{\delta^2}{\delta({E^\phi}(x))\delta({{E^x}}(x))}\right]\Psi\\
&=&\left[-({E^\phi}(x))^2+\frac{1}{4}(E^{x\prime}(x))^2+ {{E^x}}(x)E^{x\prime\prime}(x)-{{E^x}}(x)E^{x\prime}(x)({E^\phi}(x))^{-1}E^{\phi\prime}(x)\right]
\Psi. \nonumber
\eea
The form of this equation (\ref{HamEqn}) prompts an exponential ansatz for the wave functional.
\bea\label{HamEqnSoln1}
\Psi=\text{exp}\left(-k\int\mathrm{d}x\,\, {E^\phi}(x) \,\,f\left(\frac{E^{x\prime}(x)}{{E^\phi}(x)}\right)\right),
\eea
where $k$ is some constant yet to be fixed. As of now, the functional $f$ is still undetermined up to the form defined above, which will be determined as the solution of a differential equation obtained by plugging in our ansatz (\ref{HamEqnSoln1}) in (\ref{HamEqn}). The argument of the function $f$ is different here than that in (\ref{DiffeEqnSoln}) since we want to start with a relatively simple ansatz before giving the most general solution later.
\bea\label{HamEqnSoln2}
&&2k^2{{E^x}}'\dot{f}f -k^2\frac{({{E^x}}')^2}{{E^\phi}}\dot{f}^2-k^2{E^\phi}f^2 + 4k^2\frac{{{E^x}}{{E^x}}''}{{E^\phi}}\ddot{f}f -4k^2\frac{{{E^x}}{{E^x}}''{E^\phi}'}{({E^\phi})^2}\ddot{f}\dot{f} -4k^2\frac{{{E^x}}{{E^x}}'{E^\phi}'}{({E^\phi})^2}\ddot{f}f\nonumber\\
&&+4k^2\frac{{{E^x}}({{E^x}}')^2{E^\phi}'}{({E^\phi})^3}\ddot{f}\dot{f}+{E^\phi}-\frac{1}{4}\frac{({{E^x}}')^2}{{E^\phi}}-\frac{{{E^x}} {{E^x}}''}{{E^\phi}}+\frac{{{E^x}}{{E^x}}'{E^\phi}'}{({E^\phi})^2}=0,
\eea
where a `dot' on $f$ denotes a derivative with respect to its argument $\left(\frac{{{E^x}}'}{{E^\phi}}\right)$. A solution to the above differential equation turns out to be
\bea\label{HamEqnSoln3}
f\left(\frac{{{E^x}}'}{{E^\phi}}\right)=\left(\frac{{{E^x}}'}{{E^\phi}}\right)\sin^{-1}\left(\frac{{{E^x}}'}{2{E^\phi}}\right) + \sqrt{4-\left(\frac{{{E^x}}'}{{E^\phi}}\right)^2},
\eea
where we have to fix $k=\pm 1/2$.
The final solution for the wave-functional can be written as (the factor $1/2$ introduced here shall be explained later on.)
\bea\label{HamEqnSoln4}
\Psi=\text{exp}\left(-\frac{1}{2}\int\mathrm{d}x\,\, {E^\phi}(x) \,\,f\left(\frac{{{E^x}}'(x)}{{E^\phi}(x)}\right)\right),
\eea
where $f$ is given by (\ref{HamEqnSoln3}).

Of course, any of the other infinite solutions from (\ref{DiffeEqnSoln}) are equally suited to be an ansatz for (\ref{HamEqn}) with the exponential functional. However, as it turns out, the most general form for these solution (up to one arbitrary parameter) can be written as
\bea\label{HamEqnSoln7}
\Psi_{\text{c}}=\text{exp}\left(-\frac{1}{2}\int\mathrm{d}x\, {E^\phi}(x)\, f_\text{c}\left[\left(\frac{{{E^x}}'}{{E^\phi}}\right),{{E^x}}(x)\right] \right),
\eea
with
\bea\label{HamEqnSoln6}
f_\text{c}\left[\left(\frac{{{E^x}}'}{{E^\phi}}\right),{{E^x}}(x)\right]=\left(\frac{{{E^x}}'}{{E^\phi}}\right)\sin^{-1}\left[\left(\frac{{{E^x}}'/{E^\phi}}{\sqrt{4+(c/{{E^x}})^{1/2}}}\right)\right] + \sqrt{\left(4+\sqrt{\frac{c}{{{E^x}}}}\right)-\left(\frac{{{E^x}}'}{{E^\phi}}
\right)^2},
\eea
where $c=c(t)$ is some arbitrary function of time. The above solutions form a basis for the physical Hilbert space in the triad representation. Thus we have the same basis for $\mathcal{H}_{\text{phys}}$ as in \cite{Thiemann:1992jj}. To make a comparison, we need to identify ${{E^x}}$ and ${E^\phi}$ in our notation with $E^1$ and $\sqrt{E}$ respectively.\footnote{There is a factor of $\frac{1}{2}$ in front of the integral in the exponential, which is different from that in \cite{Thiemann:1992jj}. The reason for this discrepancy stems from the difference in the basic Poisson brackets in the two approaches. There is an overall factor of $\frac{1}{2}$ in \cite{Thiemann:1992jj} in both the brackets as compared to our convention. This also leads to a factor of $2$ discrepancy in the form of the argument of the function $f$.} It is also important to remember at this point that we have eliminated for one canonical pair by solving the Gauss constraint. Since we have quantized a system with fundamental canonical variables different from those in \cite{Thiemann:1992jj}, the resulting physical Hilbert space did not have to be the same in both cases. It is surprising that the Hilbert spaces in the two cases are identical since not only are the basic canonical variables very different in the two approaches but also the mathematical details followed in the quantization procedures are also different. For instance, the set of constraints chosen in \cite{Thiemann:1992jj} are linear in the momenta variables (a mathematical fact used at various steps of the construction) which is certainly not true for our case of the gravitational constraints. However since they do turn out to be the same, we want to stress that this exhibits the robustness of the quantization procedure. Canonical quantization of this midi-superspace model using real and self-dual (in the latter case, the Immirzi parameter is fixed to be $i$) Ashtekar variables seem to be consistent with each other.

We can now define physical states as in \cite{Thiemann:1992jj} as
\bea\label{PsiPhys}
\Psi_{\text{physical}}=\int_\mathbb{R} \mathrm{d}c \,g(c) \,\,\Psi_\text{c},
\eea
where $g(c)$ is a square integrable function on the real line. Although $\Psi_\text{c}$ itself is not normalizable, we can choose $g(c)$ to be a sharply peaked function around some value $c_0$ such that we can define a normalizable physical wavefunction.

Finally we have to get an inner product on this $\mathcal{H}_{\text{phys}}$. We shall employ an algebraic method to do this as described in \cite{Tate:1992hc}. This task is easier for us than in \cite{Thiemann:1992jj}, since both the triads and its conjugate extrinsic curvature components are real variables in this model. For the physical states defined above in (\ref{PsiPhys}), we can choose an ansatz for the inner product to be
\bea
\langle \Psi_{\text{physical}}|\Phi_{\text{physical}}\rangle = \int\,\mathrm{d}{{E^x}} \wedge \mathrm{d}{E^\phi}\, \mu({{E^x}}, {E^\phi})\, \bar{\Psi}_{\text{physical}}\, \Phi_{\text{physical}}.
\eea
Using the relations (\ref{star1}, \ref{star2}), we solve for the measure $\mu$. As we do not have `hybrid' canonical variables as in \cite{Thiemann:1992jj}, this turns out to give the trivial solution $\mu=\text{constant}$.\footnote{Requiring that the extrinsic curvature components are self-adjoint turns out to be sufficient to determine $\mu$.} Choosing $\mu=1$ gives us the obvious form of the inner product to be
\bea\label{ScalarProduct}
\langle \Psi_{\text{physical}}|\Phi_{\text{physical}}\rangle = \int\,\mathrm{d}{{E^x}} \wedge \mathrm{d}{E^\phi}\, \bar{\Psi}_{\text{physical}}\, \Phi_{\text{physical}}.
\eea
This form of the inner product again matches with the one given in \cite{Thiemann:1992jj}, except for a factor of $\exp(\int_\Sigma E^1\Gamma_1)$ in their notation. This is a factor absent for us both in the scalar product as well as the definition of a physical state. The reason is that $\Gamma_1 (= \eta'$ in our notation) is pure gauge and $\eta$ was eliminated by us right from the beginning by assuming that the Gauss constraint has been solved.

At this point, it is pertinent to discuss the form of the second order differential equation (\ref{HamEqnSoln2}), which leads to a one-parameter family of solutions (\ref{HamEqnSoln6}). This is not an ordinary differential equation for the function $f=f({E^x}'/E^{\phi}, E^x)$, since the co-efficients of the derivative operators are arbitrary functions of $E^x, E^{\phi}$ and their derivatives, and are not of the form of the argument of $f$. Thus there is no systematic way to find the general solution for such an equation and one has to come up with a suitable ansatz. The solution found here coincides with what has been derived in \cite{Thiemann:1992jj} up to factors of $2$ as discussed earlier.  This indeed corresponds to the correct physical solution as well since, for static configurations, this leads to the classical Schwarzschild solution. This analysis has been performed in detail in Section 6 of  \cite{Thiemann:1992jj} and is not repeated here. It is possible to construct observables in this context mimicking the exact same treatment in \cite{Thiemann:1992jj} which would lead to the remaining physical degree of freedom to be the Schwarzschild mass (after choosing the parameter $c$ to be proportional to the square of the Schwarzschild mass). This also implies that the physical wave-function obtained using these variables is a function of the Schwarzschild mass (for stationary space-times) which matches with results coming from using metric variables \cite{Kuchar:1994zk}. Although we do not repeat these calculations here again, the above-mentioned conclusions about classical states follow naturally since the basis of the physical Hilbert space (and the inner product on it) in our case turns out to be the same as that in \cite{Thiemann:1992jj}.

For the other choice of factor ordering, we can once again solve for states satisfying the diffeomorphism constraint first to be of the form
\bea\label{DiffeoSoln5}
\Psi=\Psi\left[\int\mathrm{d}x\,{{E^x}}'(x)\,f\left[{{E^x}}(x)\right]\right],
& &\Psi\left[\int\mathrm{d}x\,\left(\frac{{{E^x}}(x)}{{E^\phi}(x)}\right)'\,f\left[\frac{{{E^x}}(x)}{{E^\phi}(x)}\right]\right], \ldots
\eea
However, it can be shown that none of these states satisfy the Hamiltonian constraint, leading to a trivial solution for the physical Hilbert space. Thus to get a non-trivial $\mathcal{H}_{\text{phys}}$ we choose the `normal ordering' for the gravitational constraints.

\section{Deformed constraint algebra in the presence of holonomy corrections}
Loop Quantum Gravity is based on holonomy and flux operators which are obtained by `smearing' out the canonically conjugate connection components and densitized triad fields described above, with smooth functions \cite{9780511755682, 9780511755804}. Instead of working with quantum operators corresponding to the gravitational constraints in full LQG, we shall work in a formalism whereby the effects of various modifications shall be encoded by appropriate correction functions (for `effective' formulations using these ideas, see \cite{JR, Cailleteau:2011kr, Bojowald:2011aa}). To begin with, these functionals can depend generically on  all the canonical field variables and their spatial derivatives. However, requiring that the constraint algebra closes in an anomaly-free manner imposes certain restrictions on these functions. In fact, \textit{a priori}, it is not obvious that the quantum-corrected constraints would form a first-class system. (In case they do form a closed algebra, the quantum-corrected constraints, as generators of gauge transformations, must eliminate the same number of spurious degrees of freedom as in the classical theory.) This is sometimes referred to as the `anomaly' problem in canonical quantum gravity \cite{Nicolai:2005mc}.   We shall introduce such correction functions only in the Hamiltonian constraint since spatial diffeomorphism invariance is implemented in the full quantum theory through finite unitary transformations. There is no inifinitesimal quantum operator generating spatial diffeomorphisms in LQG, and the Hamiltonian constraint operator acts on diff-invariant states in the full theory. Thus we shall leave the diffeomorphism constraint unchanged although we introduce correction functions in the Hamiltonian constraint operator.\footnote{We are interested in possible closed algebras of constraint operators, which to a large degree is a representation-independent question. The problems with a non-existing diffeomorphism constraint operator come up when one tries to represent it on spin-network states, so clearly in a situation in which the specific representation is important. It is then justified to use an unmodified diffeomorphism operator because the flow should not be crucially different from the classical one. However, we wish to emphasize that this is indeed an assumption for our purposes, albeit justified as above.}

Different correction functions can be introduced to account for different non-perturbative quantum effects coming from LQG. Here we shall only be concerned with corrections due to the use of holonomies instead of connections in LQG. Also, as a first approximation, we shall require that the correction functions only depend on the ${{K_\phi}}$ component and not on any spatial derivatives. This, in turn, implies that we are only considering point-wise holonomy corrections coming from the angular extrinsic curvature component and not considering corrections coming from using the holonomy of the radial component $K_x$.  Corrections from the latter are more difficult to implement even in `effective' theories and should come in the form of non-local corrections, which might be expanded in a formal derivative series if possible as approached in \cite{Bojowald:2014rma}. The algebra of basic variables are thus modified as
\bea\label{algHol}
\left[\hat{K}_x(x),\hat{E}^x(y)\right]&=& 2i\hbar G\delta(x-y),\\
\left[\widehat{f\left({K}_{\phi}(x)\right)},\hat{E}^{\phi}(y)\right]&=& i\hbar G \widehat{ \left(\frac{df}{d{K_\phi}}(x)\right)}\delta(x-y).
\eea

In the following, our main aim is to show that the Dirac space-time algebra can indeed still be closed even after considering such point-wise holonomy modifications, although the structure functions get deformed in such cases (at the same level of formality). Let us start by introducing the modified Hamiltonian constraint with holonomy corrections, which are encoded by functions of the (angular) extrinsic curvature component
\bea\label{HamQH}
H[N]&=&\frac{-1}{2G}\int \mathrm{d}x \,\,N\Bigg[({{E^x}})^{-1/2}f_1({K_\phi}){E^{\phi}} + 2({{E^x}})^{1/2}K_{x}({E^{\phi}})^{-1}f_2({K_\phi}){E^{\phi}} +({{E^x}})^{-1/2}{E^{\phi}}\nonumber\\
& & - \frac{1}{4}({{E^x}})^{-1/2}(E^{x\prime})^{2}({E^{\phi}})^{-1} - ({{E^x}})^{1/2}E^{x\prime\prime}({E^{\phi}})^{-1} +({{E^x}})^{1/2}E^{x\prime}({E^{\phi}})^{-2}E^{\phi\prime}\Bigg].
\eea
We have to be more careful with the factor ordering in this case compared to (\ref{HamQ}) as is evident from the form of the second term. Classically there is no ${E^\phi}$ in this term. However, we need this (non-trivial) form of the operator ordering for a consistent constraint algebra.

The quantum correction functions $f_1({K_\phi})$ and $f_2({K_\phi})$ are not both independent but related to each other to ensure we have a closed algebra, as shown in Appendix B. Classically, $f_1({K_\phi})={K_\phi}^2$ while $f_2({K_\phi})={K_\phi}$. Although the holonomy corrections here are kept unspecified for our purposes, they can take specific form such as a periodic function of the extrinsic curvature. Of course, the modification must also be such that these functions have the correct classical limit\footnote{One such choice would be $f_1({K_\phi})=2\left(\frac{1-\cos(\gamma\delta {K_\phi})}{(\gamma\delta)^2}\right)$, for holonomies of compact groups. The $\delta$ here is related to some scale, say $l_p$, quantum gravity effects are supposed to become relevant. This gives the required classical limit and is a bounded function of the extrinsic curvature as expected.}. The diffeomorphism constraint remains unchanged as stated above
\bea\label{diffeoHQ1}
D[N^x]=\frac{1}{2G}\int\mathrm{d}x \,\,N^x\left[2K'_\phi {E^\phi} - {{E^x}}'K_x\right].
\eea
The commutator of the diffeomorphism constraint operator with itself and with the Hamiltonian constraint operator remains the same. However the commutator between two Hamiltonian constraint operators is modified as shown in Appendix B. The resulting Dirac space-time algebra is deformed as follows
\bea\label{defalgebra}
 \left[D[N^x], D[M^x]\right] &=& D[\mathcal{L}_{N^x}M^x],\\
 \left[H[N], D[N^x]\right] &=& -H[\mathcal{L}_{N^x}N],\\
 \left[H[N], H[M]\right] &=& D\left[(NM'-MN')|{{E^x}}|({E^{\phi}})^{-2}\left(\frac{d^2f_1({K_\phi})}
 {d{K_\phi}^2}\right)\right].
\eea
The structure function in the last equation is deformed by the second derivative of the function coming from holonomy corrections of LQG. Although we do not specify the exact form of the holonomy correction function, we do know that it is a bounded function of the extrinsic curvature component. In fact, generic singularity resolution in Loop Quantum Gravity models result from replacing connection components by their holonomies, which are represented by such bounded functions. Indeed if this function is bounded, then at its maximum value, the second derivative must be negative. This is interpreted sometimes as `signature-change' \cite{Mielczarek:2012pf, Cailleteau:2011kr, Bojowald:2011aa} since this now has the right sign as in Euclidean GR. This is consistent with the interpretation of the inverse of the spatial metric effectively changing sign in presence of such deformations (although the usual space-time picture is more fuzzy in the presence of such deformations \cite{Gomar:2014wta, Bojowald:2015gra} due to lack of a classical metric variable). Our main aim here is to show that, in a formal sense, there are consistent (yet, highly non-trivial) factor-ordering choices for the constraint operators even when we include some holonomy modifications from LQG. More interestingly, although the algebra of constraints is still closed, this leads to deformations of the effective structure functions in this scheme. Although this is sufficient to show the nature of the deformed Dirac space-time algebra in this setting, specific forms of these correction function must be introduced to construct a physical Hilbert space. However we do not aim to do so in this article, and leave it for later work.

\section{Conclusion}
Quantizing gravity in the full $(3+1)-$dimensional theory has been a daunting task for several years. This leads us to look towards symmetry-reduced toy models like spherically symmetric gravity to apply the general tools of canonical quantum gravity. Our main aim in
this work has been two-fold:
\begin{enumerate}                                                                                                                                                                                                                                              \item To show that the canonical quantization of this model using real-valued triads and extrinsic curvature components agree surprisingly well with that done using self-dual Ashtekar connections, as constructed in \cite{Thiemann:1992jj}. Canonically equivalent phase spaces do not necessarily generate unitarily equivalent quantum theories. It is indeed rather satisfactory to have the same physical Hilbert space in the triad representation, coming from the two different approaches, keeping in mind the different techniques that have to be employed while dealing with complex-valued and real-valued basic variables.
\item To get a consistent Dirac ordering for the gravitational constraint operators, even when some holonomy modifications from LQG are taken into account and to show how the resulting algebra differs from the standard Wheeler-DeWitt case. Several interesting new features are touched upon in this respect, namely, that not all consistent factor-ordering choices can give rise to a non-trivial physical Hilbert space. This is in accordance with the expectation that different operator orderings leads to different physical solutions. The results are consistent with results from `effective' theories \cite{JR, Cailleteau:2011kr, Bojowald:2011aa}, although our formalism goes beyond them.                                                                                                                                                                                                                                       \end{enumerate}

Building on similar methods used in this paper, future work would be focussed on constructing a similar physical Hilbert space for holonomy-corrected constraints, whereby we would have to choose a particular form for the correction functions. Another possible future extension would be to include matter degrees of freedom on such deformed algebras.

\section*{Acknowledgements}
This work was supported in part by NSF grant PHY-1307408. The author wishes to thank Umut B\"{u}y\"{u}k\c{c}am for a careful reading of the draft and is particularly grateful to Martin Bojowald for his encouragement, discussions and several suggestions on the draft. The author also thanks the referees for pointing out some very useful references.

\section*{Appendix A}
When calculating the commutators between various terms of the constraint operators, it is instructive to keep several things in mind.\footnote{For these calculations, we suppress factors of $(i\hbar)$ coming from the basic commutators to simplify the notation.} Obviously, terms that consist only of triads will commute with each other. If we focus on the commutator between two of the same constraints, i.e. $[D,D]$ or $[H,H]$, then only those terms will survive when at least one of the terms contain the spatial derivative of the triad and the other a connection component. It is so because only in that case do we get the derivative of a delta function which prevents the term from being cancelled by another identical term from the commutator. We shall see this in detail below.

Let us start by calculating the commutator between the two Hamiltonian constraints $\left[H[N],H[M]\right]$, which is the most non-trivial of all the commutators since it has a structure function on the right. The form of the Hamiltonian constraint operator is given in (\ref{HamQ}). The bracket between the first term and the last one is shown below:
\bea\label{HH1.1}
\frac{1}{4G^2}\int \mathrm{d}x \mathrm{d}y \,N(x)M(y) \left[({{E^x}}(x))^{-1/2}{E^{\phi}}(x){K_\phi}^2(x), ({{E^x}}(y))^{1/2}{{E^x}}'(y)({E^{\phi}}(y))^{-2}{E^{\phi}}'(y)' \right].
\eea
There is another identical term coming from the commutator between the last term of the first Hamiltonian constraint $H[N]$ and the first term of the second constraint $H[M]$.
\bea\label{HH1.2}
\frac{1}{4G^2}\int \mathrm{d}x \mathrm{d}y \,N(x)M(y) \left[({{E^x}}(x))^{1/2}{{E^x}}'(x)({E^{\phi}}(x))^{-2}{E^{\phi}}'(x) , ({{E^x}}(y))^{-1/2}{E^{\phi}}(y){K_\phi}^2(y) \right].
\eea
Calculating these two terms in (\ref{HH1.1}, \ref{HH1.2}), we get
\bea\label{HH1.3}
& &\frac{1}{2G}\int \mathrm{d}x \mathrm{d}y \,\,\left[N(x)M(y)\frac{d}{dy}[\delta(x-y)] - N(y)M(x)\frac{d}{dy}[\delta(y-x)]\right]\times \nonumber\\
& &\,\,\,\,\,\,\,\,\,\,\,\,\,\, \left\{({{E^x}}(x))^{-1/2}{E^\phi}(x){{E^x}}'(y)({{E^x}}(y))^{1/2}
({E^\phi}(y))^{-2}{K_\phi}(x)\right\},
\eea
where we have exchanged the dummy variables $x$ and $y$ in (\ref{HH1.2}). Now carrying out an integration by parts, and throwing away a surface term, we are left with two terms of the form
\bea\label{HH1.4}
& &\frac{1}{2G}\int \mathrm{d}x \mathrm{d}y\,\, \left[N(x)M(y) - N(y)M(x)\right]\delta(y-x)\times \nonumber\\
& &\,\,\,\,\,\,\,\,\,\,\,\,\,\, \frac{d}{dy}\left\{({{E^x}}(x))^{-1/2}{E^\phi}(x){{E^x}}'(y)({{E^x}}(y))^{1/2}({E^\phi}(y))^{-2}{K_\phi}(x)\right\}\nonumber\\
&+& \frac{1}{2G}\int \mathrm{d}x\mathrm{d}y\,\, \left(N'(y)M(x)-M'(y)N(x)\right)\delta(x-y)\times \nonumber\\
& &\,\,\,\,\,\,\,\,\,\,\,\,\,\, \left\{({{E^x}}(x))^{-1/2}{E^\phi}(x){{E^x}}'(y)({{E^x}}(y))^{1/2}({E^\phi}(y))^{-2}{K_\phi}(x)\right\}.
\eea
Obviously the first term just cancels out and we are left with the term below
\bea\label{HH1.5}
\frac{1}{2G}\int \mathrm{d}x\, \left(N'M-M'N\right)\left\{({{E^x}})'({E^\phi})^{-1}{K_\phi}\right\},
\eea
where we have suppressed the dependence of the field variables on $x$.
This also illustrates why only those terms survive in the commutator which gives rise to the derivative of the delta function, and not just the delta function itself.

Next we consider the commutator between the second and fourth terms of the Hamiltonian constraints.
\bea\label{HH2.1}
& &\frac{1}{4G^2}\int \mathrm{d}x \mathrm{d}y \,N(x)M(y) \left[2({{E^x}}(x))^{1/2}{K_\phi}(x)K_{x}(x), -\frac{1}{4}({{E^x}}(y))^{-1/2}({{E^x}}'(y))^2({E^{\phi}}(y))^{-1} \right]\nonumber\\
&=&\frac{1}{2G}\int \mathrm{d}x \mathrm{d}y\,\, N(x)M(y)\left\{({{E^x}}(x))^{1/2}({{E^x}}(y))^{-1/2}({E^\phi}(y))^{-1}{{E^x}}'(y)
{K_\phi}(x)\right\}\frac{d}{dy}[\delta(x-y)].\nonumber\\
\eea
Combining this with the corresponding bracket between the fourth and the second terms, and performing integration by parts like before, we get
\bea\label{HH2.2}
-\frac{1}{2G}\int \mathrm{d}x\, \left(N'M-M'N\right)\left\{({{E^x}})'({E^\phi})^{-1}{K_\phi}\right\}.
\eea
Thus the terms in (\ref{HH1.5}) and (\ref{HH2.2}) cancel each other.
Another calculation similar to the above comes from the commutator between the second and last terms of the Hamiltonian constraints. We employ standard commutator formulae of the form $[AB,CD]=A[B,C]D+CA[B,D]+[A,C]BD+C[A,D]B$, to calculate the term below
\bea\label{HH3.1}
& &\frac{1}{4G^2}\int \mathrm{d}x \mathrm{d}y \,N(x)M(y) \left[2({{E^x}}(x))^{1/2}{K_\phi}(x)K_{x}(x), ({{E^x}}(y))^{1/2}{{E^x}}'(y)({E^{\phi}}(y))^{-2} {E^{\phi}}'(y)\right]\nonumber\\
&=&\frac{1}{2G}\int \mathrm{d}x \mathrm{d}y\,\, N(x)M(y)\left\{({{E^x}}(x))^{1/2}({{E^x}}(y))^{1/2}({E^\phi}(y))^{-2}\right\}\times\nonumber\\ &  &\,\,\,\,\,\,\,\,\,\,\,\,\,\,\,\left\{K_x(x){{E^x}}'(y)+2{E^\phi}'(y){K_\phi}(x)
\right\}\frac{d}{dy}[\delta(x-y)].
\eea
Combining the above with the corresponding commutator between the last term of $H[N]$ and the second term of $H[M]$, we get
\bea\label{HH3.2}
\frac{1}{2G}\int \mathrm{d}x\,(N'M-M'N)\left\{{{E^x}}({E^\phi})^{-2}\right\}\left\{K_x E^{x\prime}
+2{E^\phi}'{K_\phi}\right\}.
\eea
We immediately observe that the first term above is one of the required terms in the end. Finally we must calculate the commutator between the second term of $H[N]$ and the fifth term of $H[M]$. This is a slightly different calculation from the rest since there is a second derivative on one of the triad terms:
\bea\label{HH4.1}
& &\frac{1}{4G^2}\int \mathrm{d}x \mathrm{d}y \,N(x)M(y) \left[2({{E^x}}(x))^{1/2}{K_\phi}(x)K_{x}(x), -({{E^x}}(y))^{1/2}{{E^x}}''(y)({E^{\phi}}(y))^{-1} \right]\nonumber\\
&=&-\frac{1}{G}\int \mathrm{d}x \mathrm{d}y \,N(x)M(y)\left\{({{E^x}}(x))^{1/2}({{E^x}}(y))^{1/2}({E^\phi}(y))^{-1}
{K_\phi}(x)\right\}\frac{d^2}{dy^2}[\delta(x-y)].
\eea
After performing a couple of integration by parts, we get three terms given by
\bea\label{HH4.2}
&-&\frac{1}{G}\int \mathrm{d}x \mathrm{d}y \,N(x)M''(y)\left\{({{E^x}}(x))^{1/2}({{E^x}}(y))^{1/2}({E^\phi}(y))^{-1}{K_\phi}(x)\right\}\delta(x-y)\nonumber\\
&-&\frac{2}{G}\int \mathrm{d}x \mathrm{d}y \,N(x)M'(y)\frac{d}{dy}\left\{({{E^x}}(x))^{1/2}({{E^x}}(y))^{1/2}({E^\phi}(y))^{-1}{K_\phi}(x)\right\}\delta(x-y)\nonumber\\
&-&\frac{1}{G}\int \mathrm{d}x \mathrm{d}y \,N(x)M(y)\frac{d^2}{dy^2}\left\{({{E^x}}(x))^{1/2}({{E^x}}(y))^{1/2}
({E^\phi}(y))^{-1}{K_\phi}(x)\right\}\delta(x-y).
\eea
Now if we do the same calculation for the commutator between the fifth term of $H[N]$ and the second term of $H[M]$, we are left with three similar term like above:
\bea\label{HH4.3}
& &\frac{1}{G}\int \mathrm{d}x \mathrm{d}y \,N''(y)M(x)\left\{({{E^x}}(x))^{1/2}({{E^x}}(y))^{1/2}({E^\phi}(y))^{-1}{K_\phi}(x)\right\}\delta(y-x)\nonumber\\
&+&\frac{2}{G}\int \mathrm{d}x \mathrm{d}y \,N'(y)M(x)\frac{d}{dy}\left\{({{E^x}}(x))^{1/2}({{E^x}}(y))^{1/2}({E^\phi}(y))^{-1}{K_\phi}(x)\right\}\delta(y-x)\nonumber\\
&+&\frac{1}{G}\int \mathrm{d}x \mathrm{d}y \,N(y)M(x)\frac{d^2}{dy^2}\left\{({{E^x}}(x))^{1/2}({{E^x}}(y))^{1/2}
({E^\phi}(y))^{-1}{K_\phi}(x)\right\}\delta(x-y).
\eea
The last term in (\ref{HH4.2}) is cancelled by the last term in (\ref{HH4.3}). The terms which are left over, can be combined to give
\bea\label{HH4.4}
& &\frac{1}{G}\int \mathrm{d}x\, (N'M-M'N)\left\{({E^\phi})^{-1}{{E^x}}'{K_\phi}- 2{{E^x}}({E^\phi})^{-2}{E^\phi}'{K_\phi}\right\}\nonumber\\
&+&\frac{1}{G}\int \mathrm{d}x\, (N''M-M''N)\left\{({E^\phi})^{-1}{{E^x}}{K_\phi}\right\}.
\eea
Performing integration by parts on the last term in the above equation, and combining with the other term, we finally get
\bea\label{HH4.5}
-\frac{1}{G}\int \mathrm{d}x\, (N'M-M'N)\left\{{{E^x}}({E^\phi})^{-1}K'_\phi+{{E^x}}({E^\phi})^{-2}{E^\phi}'
{K_\phi}\right\}.
\eea
We notice that the first term in (\ref{HH4.5}) cancels the second term in (\ref{HH3.2}). The terms that remain gives us
\bea
& &\left[H[N],H[M]\right]\\
&=&\frac{1}{2G}\int\mathrm{d}x\,\left[N(x)M'(x)-N'(x)M(x)\right]\left\{{{E^x}}(x)({E^\phi}(x))^{-2}\right\}\left\{2K'_\phi(x) {E^\phi}(x)-K_x(x) {{E^x}}'(x)\right\}.\nonumber
\eea
The RHS of the $[H,H]$ commutator is exactly what we require for our particular choice of the diffeomorphism constraint.

For our factor ordering choice of the gravitational constraints as above, the $[D,D]$ and the $[H,D]$ commutators can be easily shown to satisfy the required operator relations. We do not show the details of those calculations as they are similar to the above calculations and yet much simpler due to absence of any phase space functions on the RHS. In fact these two relations show that the spatial diffeomorphism algebra is a subalgebra free of structure functions, and thus forms a usual Lie algebra.

\section*{Appendix B}
For the particular ordering choice for the holonomy corrected Hamiltonian constraint operator, we show that the $[H,H]$ bracket closes into the diffeomorphism constraint, albeit with a deformed structure function. The $[D,D]$ and $[H,D]$ commutators remain unaltered.

Like before, we look at brackets between various terms of the Hamiltonian operator (\ref{HamQH}). The bracket between the first term and the last one is shown below:
\bea\label{HHH1.1}
\frac{1}{4G^2}\int \mathrm{d}x \mathrm{d}y \,N(x)M(y) \left[({{E^x}}(x))^{-1/2}f_1({K_\phi}(x)){E^{\phi}}(x), ({{E^x}}(y))^{1/2}(E^{x\prime}(y))({E^{\phi}}(y))^{-2}({E^{\phi\prime}}(y)) \right].
\eea
Combining the above with the corresponding commutator between the last term of $H[N]$ and the first term of $H[M]$, we get
\bea\label{HHH1.2}
\frac{1}{4G}\int \mathrm{d}x\, \left(N'M-M'N\right)\left\{(E^{x\prime})({E^\phi})^{-2}\dot{f}_1({K_\phi})
{E^\phi}\right\}.
\eea
The dot on $f_1({K_\phi})$ denotes a derivative with respect to ${K_\phi}$. A similar calculation between the second and fourth terms of the Hamiltonian constraint yields
\bea\label{HHH2.1}
\frac{1}{2G}\int \mathrm{d}x\, \left(N'M-M'N\right)\left\{(E^{x\prime})({E^\phi})^{-2}f_2({K_\phi}){E^\phi}\right\}.
\eea
From the previous calculation we know that these two terms (\ref{HHH1.2}) and (\ref{HHH2.1}) must cancel each other in order to make the algebra close. This imposes a relation between the so-far unconstrained correction functions of the form
\bea\label{corrfunc}
f_2({K_\phi})=\frac{1}{2}\frac{\mathrm{d}f_1({K_\phi})}{\mathrm{d}{K_\phi}}.
\eea
This has the same form as what is obtained from `effective' theories of Loop Quantum Gravity \cite{JR}.

The bracket between the second and the fifth terms gives rise to terms of the form
\bea\label{HHH3.1}
& &\frac{1}{G}\int \mathrm{d}x\, (N'M-M'N)\left\{E^{x\prime}({E^\phi})^{-2}f_2({K_\phi}){E^\phi}- 2E^{\phi\prime}{{E^x}}({E^\phi})^{-3}f_2({K_\phi}){E^\phi}\right\}\nonumber\\
&+&\frac{1}{G}\int \mathrm{d}x\, (N''M-M''N)\left\{{{E^x}}({E^\phi})^{-2}f_2({K_\phi}){E^\phi}\right\}.
\eea
Performing integration by parts on the last term, we get
\bea\label{HHH3.2}
-\frac{1}{G}\int \mathrm{d}x\, (N'M-M'N)\left\{{{E^x}}({E^\phi})^{-2}\dot{f}_2({K_\phi})K'_\phi {E^\phi} +{{E^x}}({E^\phi})^{-2}f_2({K_\phi})E^{\phi\prime}\right\}.
\eea
Finally, the bracket between the second and the last term gives
\bea\label{HHH4.1}
\frac{1}{2G}\int \mathrm{d}x\,(N'M-M'N)\left\{{{E^x}}({E^\phi})^{-2}\right\}\left\{\dot{f}_2(K_x) E^{x\prime}K_x+2f_2({K_\phi})E^{\phi\prime}\right\}.
\eea
The second term of (\ref{HHH4.1}) cancels the second term of (\ref{HHH3.2}) and the final form of the commutator is
\bea
\left[H[N],H[M]\right]
&=&\frac{1}{2G}\int\mathrm{d}x\,\left[N(x)M'(x)-N'(x)M(x)\right]
\left\{{{E^x}}(x)({E^\phi}(x))^{-2}\left(\frac{d^2f_1({K_\phi})}{d{K_\phi}^2}\right)\right\}\nonumber\\
& &\,\,\,\,\,\,\,\,\,\,\,\,\,\,\,\,\,\,\,\,\,\,\,\,\,\,\,\,\,\,\,\,\,\,\,\,\,\,\,\,\,\,\,\left\{2K'_\phi(x) {E^\phi}(x)- E^{x\prime}(x) K_x(x)\right\}.
\eea
The term on the RHS is the diffeomorphism constraint as in (\ref{diffeoHQ1}). We have replaced the correction function $f_2$ in terms of the function $f_1$, whereby the structure function now is no longer just the inverse of the spatial metric. It is now multiplied by the second derivative of the correction function (which replaces the extrinsic curvature component with its corresponding holonomy function).

\bibliographystyle{ieeetr}
\bibliography{refs}

\end{document}